\documentclass[twocolumn,superscriptaddress,amsmath,amssymb,aps,longbibliography,nofootinbib,prl]{revtex4-1}
\usepackage[colorlinks=true,urlcolor=blue,citecolor=blue,linkcolor=blue]{hyperref}
\usepackage{txfonts}
\usepackage{graphicx}
\usepackage{bm}
\usepackage{booktabs}
\usepackage[dvipsnames]{xcolor}
\usepackage{braket}
\usepackage{dcolumn} 

\renewcommand{\vec}[1]{\bm{#1}}

\newcommand{\Eq}[1]{Eq.~(\ref{#1})}
\newcommand{\Fig}[1]{Fig.~\ref{#1}}

\begin{document}

\title{Accurate and thermodynamically consistent hydrogen equation of state \\for planetary modeling with flow matching}

\author{Hao Xie}
\email{qwexiehao@gmail.com}
\affiliation{Department of Astrophysics, University of Zürich, Winterthurerstrasse 190, 8057 Zürich, Switzerland}

\author{Saburo Howard}
\affiliation{Department of Astrophysics, University of Zürich, Winterthurerstrasse 190, 8057 Zürich, Switzerland}

\author{Guglielmo Mazzola}
\email{guglielmo.mazzola@uzh.ch}
\affiliation{Department of Astrophysics, University of Zürich, Winterthurerstrasse 190, 8057 Zürich, Switzerland}

\date{\today}

\begin{abstract}
Accurate determination of the equation of state of dense hydrogen is essential for understanding gas giants.
Currently, there is still no consensus on methods for calculating its entropy, which play a fundamental role and can result in qualitatively different predictions for Jupiter's interior.
Here, we investigate various aspects of entropy calculation for dense hydrogen based on \emph{ab initio} molecular dynamics simulations. Specifically, we employ the recently developed flow matching method to validate the accuracy of the traditional thermodynamic integration approach. We then clearly identify pitfalls in previous attempts and propose a reliable framework for constructing the hydrogen equation of state, which is accurate and thermodynamically consistent across a wide range of temperature and pressure conditions. This allows us to conclusively address the long-standing discrepancies in Jupiter's adiabat among earlier studies, demonstrating the potential of our approach for providing reliable equations of state of diverse materials.
\end{abstract}

\maketitle

\textit{Introduction.} The equation of state (EOS) of dense hydrogen (H) is one of the core ingredients to model the internal structure of giant gaseous planets, such as Jupiter and Saturn~\cite{Helled2020}. 
With the wealth of data from recent spacecraft missions, \textit{Galileo}~\cite{Johnson1992} and \textit{Juno}~\cite{Bolton2017}, which measured Jupiter's atmospheric composition and gravitational field, the lack of accurate knowledge of the thermodynamic properties of hydrogen is now considered as one of the major roadblocks in planetary science~\cite{Miguel2016, Debras_2019, Mazevet2022, howard2023}. As parameterized planetary models already require several hypotheses~\cite{Debras_2019, Nettelmann_2021, miguel2022, Militzer_2022}, an uncontested EOS is needed to advance the field.

Since high-pressure experiments are difficult to perform and still cannot yield precise results, one usually relies instead on numerical simulations, typically \emph{ab initio} molecular dynamics (MD) based on electron structure methods such as density functional theory (DFT) or quantum Monte Carlo (QMC). Up to now, there have been several widely used \emph{ab initio} EOSs for planetary science applications, including the one by Chabrier, Mazevet and Soubiran (CMS19)~\cite{Chabrier_2019}, by Militzer and Hubbard (MH13)~\cite{Militzer_2013}, and the Rostock EOS (REOS2~\cite{Nettelmann_2012} and REOS3~\cite{Becker_2014}) of various versions. All of them are derived from DFT-MD simulations using the exchange-correlation functional of Perdew, Burke and Ernzerhof (PBE)~\cite{PhysRevLett.77.3865}. Despite this common origin, however, they can make significant differences in the resulting Jupiter model, such as the location of adiabat and the predicted size of core mass~\cite{Miguel2016, howard2023}.

The key factor behind these discrepancies of planetary models is the calculation of entropy, which has been recognized and discussed to some extent in several studies~\cite{Militzer_2013, Miguel2016, Mazevet2022}. Unlike observables such as energy $E$ and pressure $p$, the Helmholtz free energy $F$ and entropy $S = (E-F)/T$ are not directly accessible from \emph{ab initio} simulations and often rely on extra procedures called thermodynamic integration (TI). One common implementation of this technique involves interpolating the available data to create a continuous path on the temperature-density phase diagram~\cite{Nettelmann_2012, Miguel2016}. From basic statistical mechanics we have
\begin{equation}
    \left( \frac{\partial}{\partial T} \frac{F}{T} \right)_\rho = -\frac{E}{T^2}, \quad
    \left( \frac{\partial}{\partial \rho} \frac{F}{T} \right)_T = \frac{p}{\rho^2 T}.
    \label{eq: ti}
\end{equation}
These two equations correspond to integrating the energy (pressure) of the system along isochores (isotherms), respectively. However, this approach has been criticized~\cite{Militzer_2013} because, due to limited number of MD simulation points and inevitable statistical uncertainties in practice, the interpolation procedure may introduce systematic errors that are difficult to detect.

An alternative TI scheme is called the coupling constant integration (CCI)~\cite{10.1063/1.1749657, PhysRevB.57.8223} or Hamiltonian thermodynamic integration~\cite{frenkel2001understanding}. By connecting the target state to an artificial Hamiltonian with known thermodynamic properties, this method further allows for the calculation of \emph{absolute} entropy. However, it is also much more expensive due to the need to perform additional MD simulations and therefore only suitable for investigating a small region of the phase diagram~\cite{PhysRevE.81.021202, Militzer_2013}. Besides, one has to carefully choose relevant parameters in the artificial system and MD simulations to ensure well-converged results~\cite{PhysRevB.109.174107}, and the errors associated with discreteness and statistical fluctuation of the simulation points still remain.

The uncertainties involved in entropy calculation are further exacerbated by the fact that the EOS used for planetary modeling is usually composed of several parts, each based on different theoretical approach and has different phase region of validity. For example, although DFT-MD can provide a good description of electron correlations at intermediate temperatures and densities, especially near pressure-induced dissociation, it will quickly become inefficient at lower densities $\rho \lesssim 0.2 \, \textrm{g}/\textrm{cm}^3$. For such a region deep within the molecular phase, empirical chemical models, such as the well-known Saumon–Chabrier–van Horn (SCvH) EOS~\cite{1995ApJS...99..713S} are generally believed to be accurate and reliable. In practice, the data from these different methods need to be connected in some way. This, however, could possibly introduce critical errors, as we will demonstrate below.

In this Letter, we focus on the crucial issues of entropy calculation outlined above at the DFT-PBE level of theory, aiming to resolve long-standing disagreements among various Jupiter model predictions, which are highly sensitive to small changes in entropy~\cite{Militzer_2013, Miguel2016, Miguel2018erratum, Mazevet2022}. The work is structured as follows: First, we calculate the free energy and entropy in the DFT-MD region using a recently developed method called \emph{flow matching}~\cite{albergo2023building, lipman2023flow, liu2023flow}. As discussed below, such an independent approach is highly desired to benchmark the accuracy of TI and identify crucial deficiencies in previous works. Second, (and more importantly,) we devise a simple and efficient protocol to construct an EOS that can achieve excellent thermodynamic consistency \emph{across various theories} and a broad range of conditions. Our procedure is primarily designed to provide a reliable and uncontested input for planetary modeling, thereby forming a solid foundation for future model refinements against spacecraft measurements and further applicability to other materials~\cite{Helled2020, Debras_2019}.



\textit{Flow matching.} Consider any two given phase points $(T_0, \rho_0)$ and $(T_1, \rho_1)$ of the hydrogen system. When calculating their free energy difference using TI, one needs to create a continuous path between them by interpolation. On the other hand, flow matching is built on the framework of targeted free energy perturbation (TFEP)~\cite{PhysRevE.65.046122, PhysRevE.79.011113}, which solely relies on information about the two end states 0 and 1; see the Eqs.~(\ref{eq: TFEP}) and (\ref{eq: forward and reverse work}) below. As a result, it can avoid any errors arising from a manual interpolation procedure and thus serves as a valuable benchmark for TI.

More specifically, within the framework of TFEP, the free energy difference (scaled by temperature) between the two states can be written as the following estimators:
\begin{equation}
    \beta_1 F_1 - \beta_0 F_0 
    = -\ln \mathop{\mathbb{E}}_{\vec{x} \sim p_0(\vec{x})} \left[ e^{-\Phi_
    \rightarrow(\vec{x})} \right]
    = \ln \mathop{\mathbb{E}}_{\vec{x} \sim p_1(\vec{x})} \left[ e^{\Phi_\leftarrow(\vec{x})} \right].
    \label{eq: TFEP}
\end{equation}
Here, $\vec{x} \equiv (\vec{r}_1, \dots, \vec{r}_N)$ denotes a configuration sample of the $N$ proton coordinates, and $p_i(\vec{x}) \propto e^{-\beta_i E_i(\vec{x})}$ $(i = 0, 1)$ are the corresponding Boltzmann distributions, with $\beta_i = 1/k_B T_i$, $k_B$ being the Boltzmann constant. The quantities
\begin{subequations}
    \begin{align}
        \Phi_\rightarrow(\vec{x}) &= \beta_1 E_1(f(\vec{x})) - \beta_0 E_0(\vec{x}) - \ln \left| \det\left( \frac{\partial f(\vec{x})}{\partial \vec{x}} \right) \right|, \\
        \Phi_\leftarrow(\vec{x}) &= \beta_1 E_1(\vec{x}) - \beta_0 E_0(f^{-1}(\vec{x})) + \ln \left| \det\left( \frac{\partial f^{-1}(\vec{x})}{\partial \vec{x}} \right) \right|
    \end{align}
    \label{eq: forward and reverse work}
\end{subequations}
are known as the \emph{forward} and \emph{reverse} work, respectively~\cite{SM}. Note that the above equations hold in principle for \emph{any invertible transformation} $f$ that acts on the configuration space. In particular, when $f$ is the identity map, Eqs.~(\ref{eq: TFEP}) and (\ref{eq: forward and reverse work}) reduce to the usual formulas for importance sampling. Moreover, we note that by simply selecting one of the states to have known thermodynamic properties, it is also fairly straightforward to compute the \emph{absolute} free energy and entropy of the system.

In practice, due to the limited number of samples, \Eq{eq: TFEP} can be efficiently estimated  with low variance only for some carefully designed transformation $f$, which should bring the two states of interest close enough to ensure sufficient overlap of their probability densities. In this work, we achieve this goal by representing $f$ as a class of generative neural network called normalizing flow~\cite{Papamakarios2021} and training the network using the flow matching technique~\cite{albergo2023building, lipman2023flow, liu2023flow}. Note that unlike the demonstrative example shown in Ref.~\cite{Zhao_2023}, here the calculations are based on DFT-MD simulations and thus have \emph{ab initio} accuracy. A more detailed introduction to the relevant background is provided in the Supplemental Material~\cite{SM}.

\textit{Entropy from ab initio data.} We employ the flow matching method to study the \emph{ab initio} portion of the phase diagram with $3000 \leqslant T \leqslant 8000 \,$K and $0.3 \leqslant \rho \leqslant 1.6 \, \textrm{g}/\textrm{cm}^3$. Specifically, we repeatedly train a transformation $f$ and use \Eq{eq: TFEP} to compute the entropy difference for each neighboring pair of states on a $(T, \rho)$ grid; see Supplemental Material~\cite{SM} for an example. We also calculate the \emph{absolute} entropy of a reference state within the same framework and then shift the entropy of other states rigidly. In practice, we choose the reference point at $T = 5000 \, \textrm{K}$ and $\rho = 1.4 \, \textrm{g}/\textrm{cm}^3$~\cite{SM}, where the computed value $0.0494(1) \, \textrm{MJ/kg/K}$ of the absolute entropy is in excellent agreement with those reported by Morales \emph{et al.}~\cite{PhysRevE.81.021202} based on coupled electron-ion Monte Carlo (CEIMC) simulations.

\begin{figure}[t]
    \centering
    \includegraphics[width=\columnwidth]{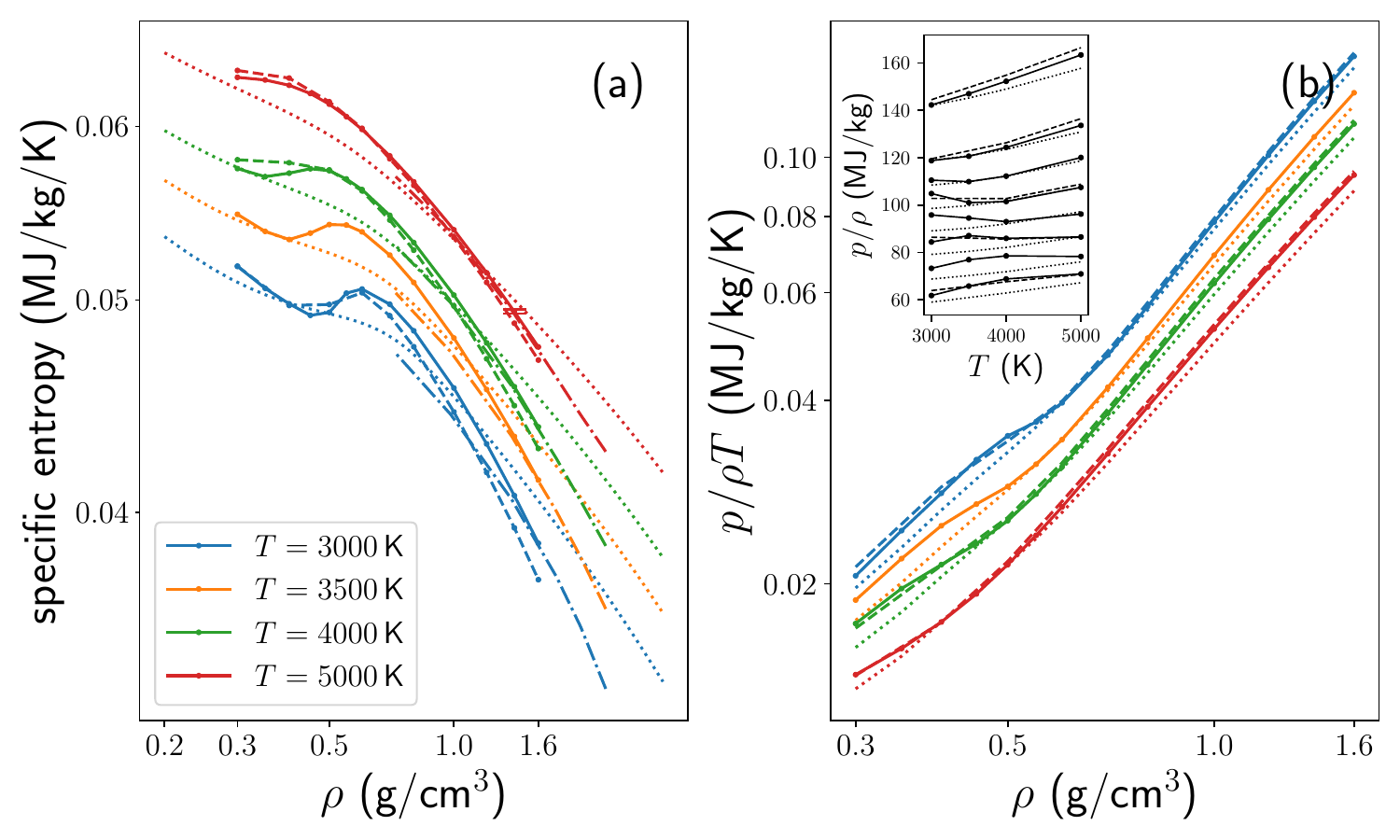}
    \caption{(a) Specific entropy $s$ and (b) pressure $p$ (scaled by $\rho T$) as a function of density $\rho$ for several isotherms. Solid line: the present work using the flow matching method; dotted line: CMS19~\cite{Chabrier_2019}; dashed line: REOS3~\cite{Becker_2014}; dash-dotted line: the entropy reported by Morales \emph{et al.}~\cite{PhysRevE.81.021202} based on CEIMC simulations. Note that the REOS3 entropy has been globally \emph{shifted} to align with our results; see text for details. The inset of (b) presents the same pressure data by selected isochores, with density ranging from $0.3$ (bottom) to $0.7 \, \textrm{g}/\textrm{cm}^3$ (top).
    }
    \label{fig: entropy_pressure}
\end{figure}

Figure~\ref{fig: entropy_pressure} shows our results (solid line) for several isotherms of the specific entropy $s$ and pressure $p$, together with two widely used hydrogen EOSs in past literature: CMS19~\cite{Chabrier_2019} (dotted line) and REOS3~\cite{Becker_2014} (dashed line). For convenience, we also include an error bar of absolute entropy at the reference point and the CEIMC entropy of Morales \emph{et al.}~\cite{PhysRevE.81.021202} (dash-dotted line) at relatively high densities. Note that the REOS3 entropy is calculated by us using TI (see \Eq{eq: ti}), and, for the sake of visualization, we have introduced a global entropy offset to make it match our data at the bottom-left phase point, i.e., $T = 3000 \, \textrm{K}$ and $\rho = 0.3 \, \textrm{g}/\textrm{cm}^3$. We do not compare with the MH13 EOS~\cite{Militzer_2013} since it was derived directly for a hydrogen-helium mixture, rather than the pure hydrogen system studied here.

One striking feature of \Fig{fig: entropy_pressure}(a) is the nonmonotonic behavior of entropy at low temperatures in both our results and REOS3, which is qualitatively different from CMS19. Moreover, CMS19 does not feature the typical slope change of the $p - \rho$ relation, as shown in \Fig{fig: entropy_pressure}(b), which is emanating from the critical point of the liquid-liquid phase transition between molecular and atomic phase~\cite{lorenzen2010, PhysRevB.100.134109}. This indicates that although derived from DFT-PBE data in the phase region considered here~\cite{Chabrier_2019, PhysRevB.83.094101, PhysRevB.77.184201, PhysRevE.81.021202}, the CMS19 EOS 
does not closely capture the pressure-induced molecular dissociation, which could have a notable impact on Jupiter models~\cite{SM}.

According to the thermodynamic relation~\cite{1995ApJS...99..713S} $(\partial s / \partial \rho)_T = - \rho^{-2} (\partial p / \partial T)_\rho$, the region of ``abnormally" \emph{increasing} entropy in \Fig{fig: entropy_pressure}(a) corresponds precisely to that of \emph{decreasing} pressure with temperature, as illustrated by the \emph{isochores} in the inset of \Fig{fig: entropy_pressure}(b). 
Physically, they are just different manifestations of the molecular dissociation effects: The attractive force from broken molecular bonds can dominate the repulsion from kinetic motion, while the larger number of configurations involving only individual H atoms (compared to those with bound H$_2$-like pairs) can compensate for the smaller occupied volume caused by increasing density.


\textit{Benchmarking thermodynamic integration.}
We use the flow matching entropy in the \emph{ab initio} region to benchmark the TI approach, which simply amounts to integrating \Eq{eq: ti} using our pressure and energy data. We find that the TI entropies are virtually indistinguishable from the solid lines in \Fig{fig: entropy_pressure}(a). Besides, they do not depend on the choice of integration path and hence show very good thermodynamic consistency. Our observations indicate that TI performs quite well for a discrete grid with spacings on the order of $1000 \,$K and $0.1 \, \textrm{g}/\textrm{cm}^3$. When combined with the absolute entropy calculation on only a \emph{single} reference point, as done above, we can then achieve the same level of accuracy as the ``full" CCI approach in Ref.~\cite{Militzer_2013} but with far less computational costs.

\begin{figure}[t]
    \centering
    \includegraphics[width=\columnwidth]{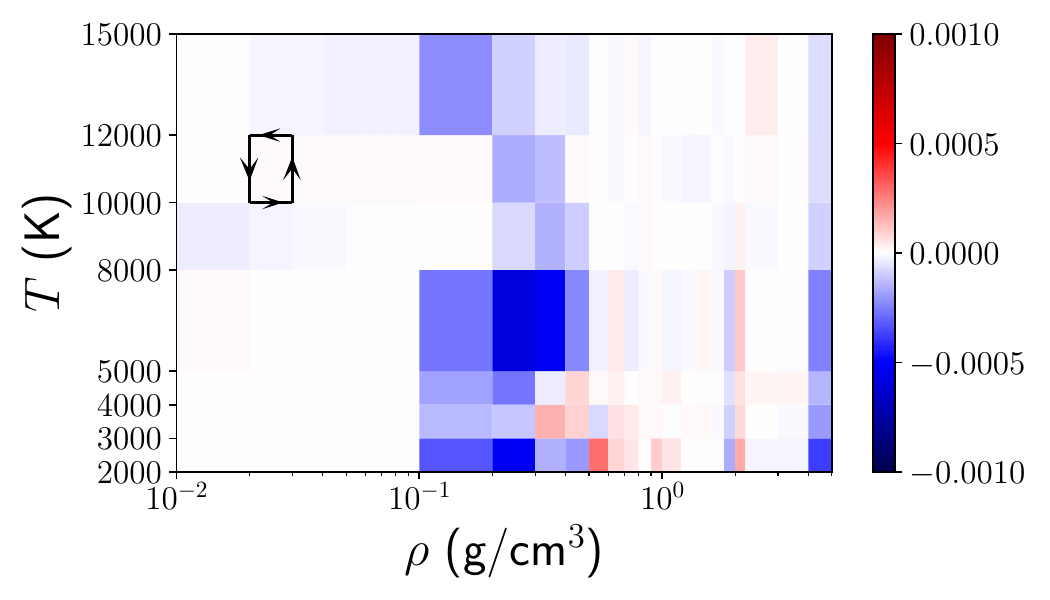}
    \caption{The line integral $\oint d(F/T)$ (in units of MJ/kg/K) computed along each closed local square loop of the REOS3 data. A schematic example of the loop is shown on the top-left corner.
    }
    \label{fig: flux_REOS3}
\end{figure}

Note that, compared to REOS3, we have selected a slightly denser grid of simulation points between $0.3$ and $0.6 \, \textrm{g}/\textrm{cm}^3$ to improve the characterization of molecular dissociation. Nevertheless, the ``discreteness" error of TI in the \emph{ab initio} region of REOS3 is still under control, and the resulting entropy remains generally consistent with our data, as shown by the dashed lines in \Fig{fig: entropy_pressure}(a). However, TI becomes truly problematic when we further extend the integration path to touch other regions based on inconsistent theories. We can quantitatively illustrate such inconsistency by performing the line integral $\oint d(F/T)$ along each closed local square loop of the tabular data, which should, in principle, vanish everywhere. However, for REOS3, we found large deviation from this ideal behavior between the \emph{ab initio} region and the chemical model at lower densities~\cite{10.1063/1.1486210, 10.1002/ctpp.200710049}, as shown in \Fig{fig: flux_REOS3}. This implies that the resulting entropy will be \emph{ambiguously} defined and strongly depend on the choice of integration path and initial reference point; see Supplemental Material~\cite{SM} for more details. One can then identify an entropy error of approximately $10\%$~\cite{SM}, which yields qualitatively large differences on planetary models derived in the past using this method~\cite{Nettelmann_2012, Miguel2016}.

\textit{Thermodynamically consistent construction.}
An inherent difficulty of the ``single TI over the entire phase diagram" approach stems from the fact that energy and pressure are different partial derivatives of the same free energy function and hence \emph{not independent}. Consequently, it turns out to be highly challenging to simultaneously interpolate \emph{both} energy and pressure data produced by different theories \emph{in a consistent way}, as demonstrated in \Fig{fig: flux_REOS3} for the case of REOS3. A better approach is to perform TI on each component (i.e., region where the same theory applies) of the EOS \emph{separately} and then interpolate the resulting \emph{free energy} function~\cite{SM}. In this way, one can derive the pressure and energy near the interpolation boundaries by taking partial derivatives, which are naturally consistent by construction. Another advantage is that one can easily incorporate accurate values of the \emph{absolute} entropy into each separate region. This ensures that any residual interpolation errors that one can make at the boundaries remain localized and do not propagate elsewhere. 

\begin{figure}[t]
    \centering
    \includegraphics[width=\columnwidth]{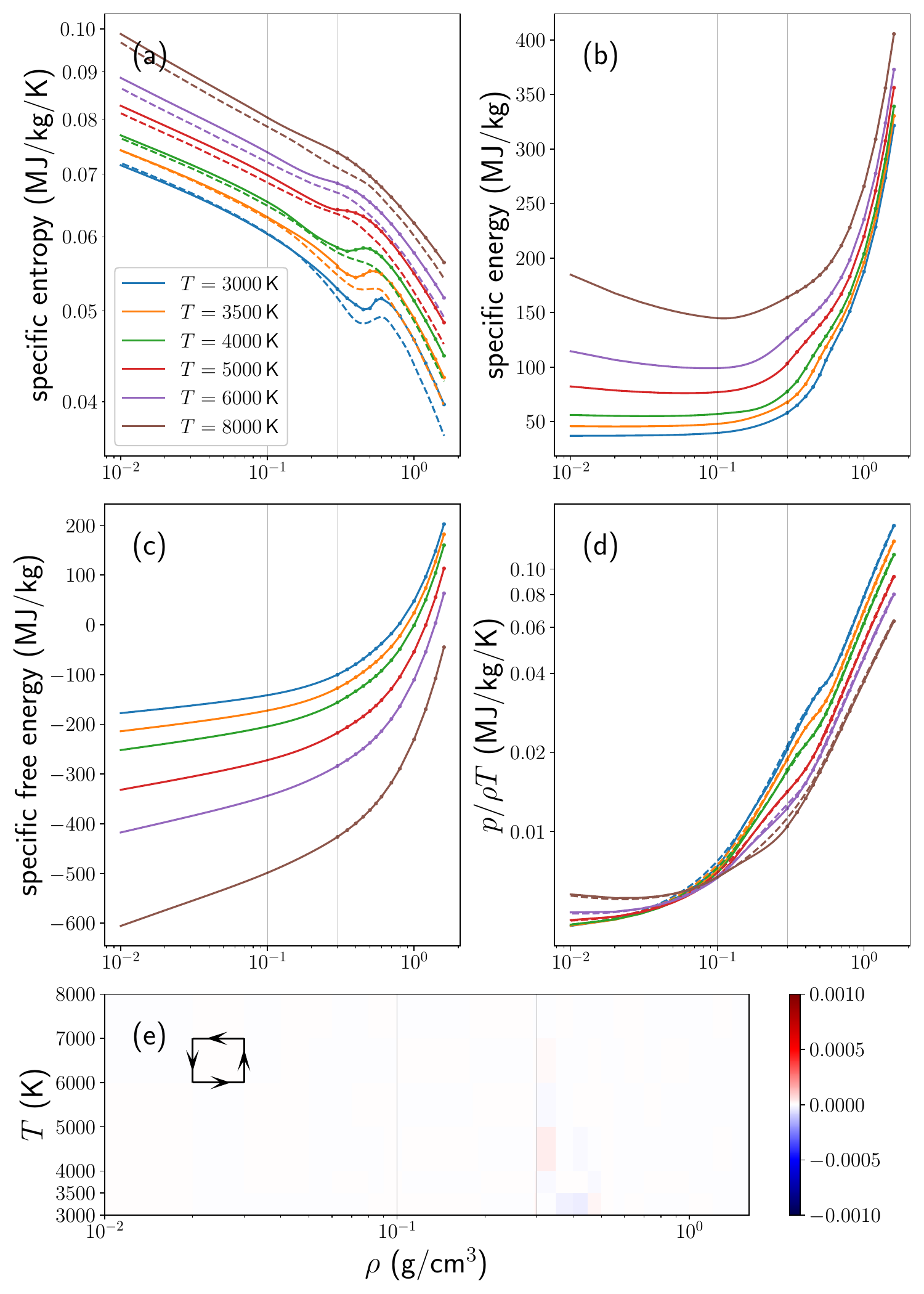}
    \caption{Several (a) specific entropy, (b) specific energy, (c) specific free energy, and (d) pressure (scaled by $\rho T$) isotherms of our final EOS (solid lines), which consist of the \emph{ab initio} data at high densities, the SCvH EOS at low densities, and an interpolation region between them, as indicated by the vertical lines.
    The dashed lines are the entropy and pressure isotherms of Miguel \emph{et al.}~\cite{Miguel2016}, obtained by performing a single TI over the entire phase region of REOS3. Note that we have added an extra entropy shift of $-0.0057 \, \textrm{MJ/kg/K}$ to both the dashed lines and SCvH part of the solid lines in panel (a); see text for details. Panel (e) shows the local loop integral values $\oint d(F/T)$ of our EOS, similar to \Fig{fig: flux_REOS3} for the case of REOS3.
    }
    \label{fig: final_eos}
\end{figure}

Figures~\ref{fig: final_eos}(a-d) show our final EOS (solid lines), which consists of the \emph{ab initio} DFT-MD part at $\rho \geqslant 0.3 \, \textrm{g}/\textrm{cm}^3$ (only above $3000 \,$K), the SCvH EOS at $\rho \leqslant 0.1 \, \textrm{g}/\textrm{cm}^3$ (covering a broader temperature range down to $\sim 100 \,$K), and an interpolation region between them. For comparison, we also include the entropy and pressure isotherms of Miguel \emph{et al.}~\cite{Miguel2016} (dashed lines), which are obtained by a single TI over the entire phase region of REOS3. Note that the new interpolation procedure depends very sensitively on the entropy difference between SCvH and the \emph{ab initio} data~\cite{SM}. In particular, a high-quality interpolation can only be achieved after globally subtracting a constant of $0.0057 \, \textrm{MJ/kg/K}$ from the originally reported SCvH entropy. Remarkably, this observation aligns precisely with various independent numerical~\cite{Chabrier_2019} and experimental references~\cite{cox1989codata, Lemmon_Thermophysical_Properties}. There, the entropy constant was recognized as $k_B \ln 2 / m_\textrm{p}$, which accounts for the contribution of proton spin degrees of freedom that are implicitly neglected on the DFT-MD side~\cite{private_comm}; see Supplemental Material~\cite{SM} for details. This strongly indicates that our new protocol is indeed the very approach to construct accurate and uncontested hydrogen EOS for planetary modeling. Figure~\ref{fig: final_eos}(e) also shows the local loop integral values $\oint d(F/T)$ of our final EOS; compared to \Fig{fig: flux_REOS3}, one clearly sees that the new protocol can indeed yield much better thermodynamic consistency over the entire phase diagram, as discussed above.


\textit{Jupiter's adiabat.}
We demonstrate the utility of the newly constructed hydrogen EOS by performing a preliminary calculation of Jupiter's adiabat. We assume a homogeneous model with no compact core and set a helium mass fraction of $Y = 0.245$ to ensure a fair comparison with previous works~\cite{Miguel2016, Miguel2018erratum, Militzer_2013}. We adopt a simple linear mixing approximation for the entropy of mixtures, using the REOS3 EOS of helium derived by Miguel \emph{et al.} in Ref.~\cite{Miguel2018erratum}. 

\begin{figure}[t]
    \centering
    \includegraphics[width=\columnwidth]{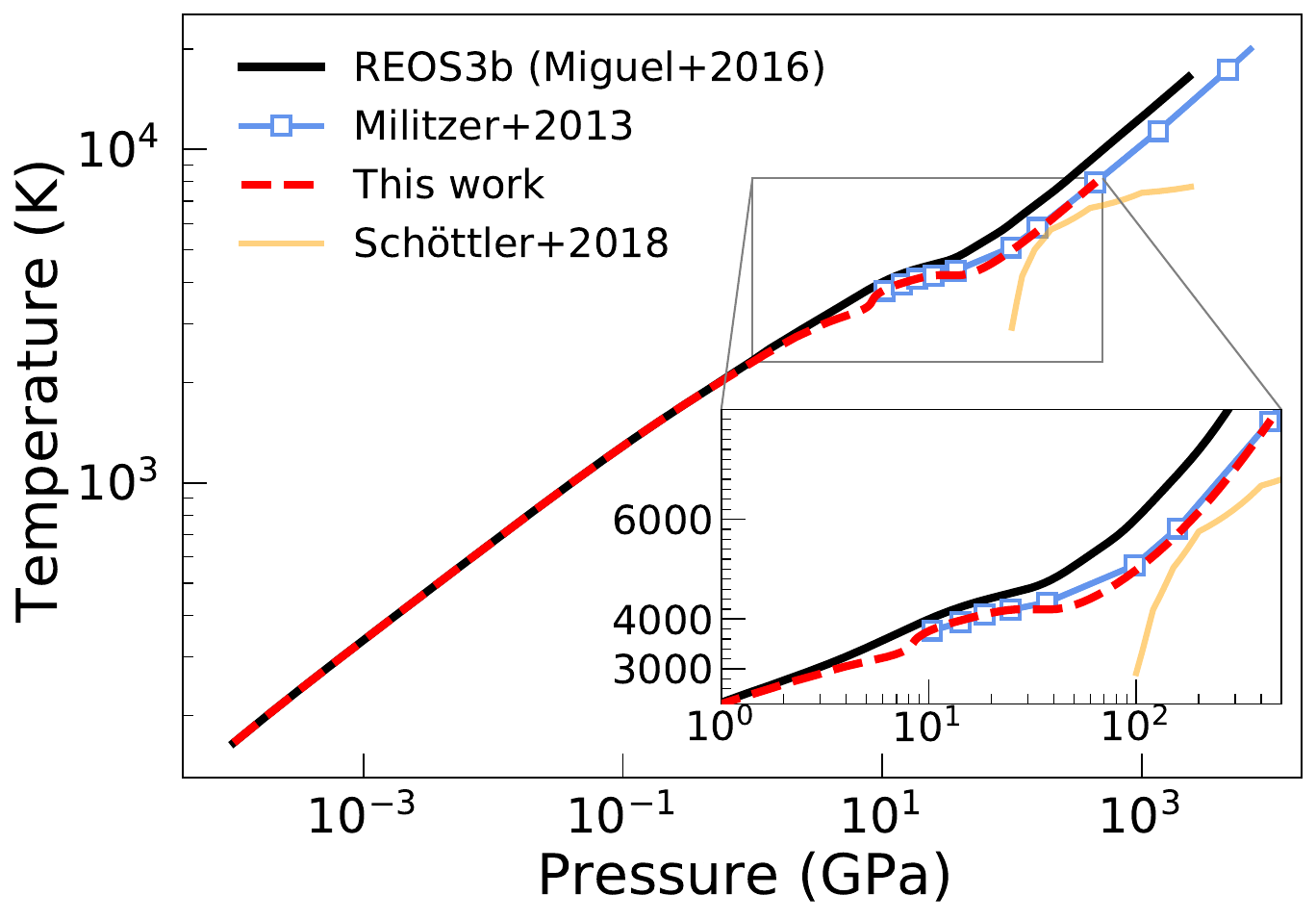}
    \caption{Jupiter's adiabats obtained from different EOSs. See text for a detailed comparison between our result (red dashed line) and those derived from the REOS3 (black line)~\cite{Miguel2018erratum} and MH13 EOS (blue line)~\cite{Militzer_2013}. The yellow line shows the hydrogen-helium demixing curve for a protosolar mixture from Ref.~\cite{PhysRevLett.120.115703}.
    }
    \label{fig:saburo}
\end{figure}

Figure~\ref{fig:saburo} shows our result for the adiabat (red dashed line) together with those derived from the REOS3 (black line)~\cite{Miguel2018erratum} and MH13 EOS (blue line)~\cite{Militzer_2013}, respectively. For a comparison also with CMS19, see the Supplemental Material~\cite{SM}. Note that the hydrogen EOS used in the REOS3 adiabat was produced by Miguel \emph{et al.}~\cite{Miguel2016, Miguel2018erratum} using a single TI over the entire phase region, as already shown by the dashed lines in \Fig{fig: final_eos}. Our result clearly agrees well with the REOS3 adiabat up to about 1GPa, which is as expected since both the underlying hydrogen EOSs are derived at low pressures from the same SCvH model. However, at higher pressures, our adiabat begins to diverge from the REOS3 line and becomes cooler. This is a direct consequence of the systematically lower REOS3 entropy in the \emph{ab initio} region in \Fig{fig: final_eos}(a), which is in turn caused by the thermodynamic inconsistency shown in \Fig{fig: flux_REOS3}~\cite{SM}. Notice that such a cooler adiabat passes very close to the miscibility boundary for hydrogen-helium mixtures, shown by the yellow line in \Fig{fig:saburo}, albeit derived from DFT-MD simulations using an exchange-correlation functional different from PBE~\cite{PhysRevLett.120.115703}. This suggests an increased likelihood of phase separation between hydrogen and helium in Jupiter’s deep interior, causing the formation of so-called helium rain~\cite{Helled2020, PhysRevB.75.024206, Chang2024}.


On the other hand, our adiabat at high pressures is apparently closer to the MH13 prediction, as shown by the blue line in \Fig{fig:saburo}. Note that the MH13 EOS is derived directly from \emph{ab initio} MD simulations of a hydrogen-helium mixture and does not rely on the linear mixing approximation. Overall, we show that, at the DFT-PBE level of theory, the long-standing discrepancy between the Jupiter models of REOS3 and MH13 primarily stems from a problematic TI calculation of the REOS3 entropy that is not everywhere thermodynamically consistent. However, according to our earlier benchmark using flow matching, this definitely does \emph{not} imply that TI is inherently inferior to the CCI approach for a finite number of simulation points, as speculated in Ref.~\cite{Militzer_2013}. Another observation from \Fig{fig:saburo} is that the MH13 adiabat~\cite{Militzer_2013} was derived only for the \emph{ab initio} region at high pressures. In fact, Ref.~\cite{Militzer_2013} did not really attempt a thermodynamic consistent interpolation with other regions to construct a complete EOS, which, however, is definitely needed for a thorough study of planetary models.

\textit{Summary and outlook.}
In this Letter, we closely examine the entropy calculation of dense hydrogen in the \emph{ab initio} region by benchmarking TI with an independent flow matching method. Building on this, we then construct a new hydrogen EOS that is both accurate and thermodynamically consistent over several orders of magnitude of pressure (roughly 1bar $\sim$ 700GPa). Remarkably, the interpolation quality at the boundaries serves as a highly sensitive and stringent ``detector" for the accuracy of the theories on both sides~\cite{SM}. Such a ``self-correcting” mechanism of the new protocol is exactly what we need to ensure steady convergence toward a truly reliable and conclusive hydrogen EOS for planetary modeling.

We conclusively address the long-standing issue of entropy disagreements among various popular EOSs used in planetary modeling~\cite{Chabrier_2019, Militzer_2013, Becker_2014}, all of which are derived from the same DFT-PBE level of electronic structure theory.
To further improve the hydrogen EOS and potentially reconcile models with observational data~\cite{Debras_2019}, one can upgrade the description of electronic correlation from DFT-PBE to a higher-level theory~\cite{doi:10.1073/pnas.1603853113, PhysRevLett.120.025701, 10.1063/5.0005037, PhysRevLett.131.126501}.
We also note that as a generic tool, flow matching can also be used to study other rich phases of hydrogen~\cite{doi:10.1073/pnas.1007309107, PhysRevLett.130.076102} or other interesting condensed matter systems~\cite{Chang2024, Wang2023}, where accurate calculation of free energy plays a significant role.

\begin{acknowledgments}
We acknowledge useful discussions with Lei Wang, Ravit Helled, Didier Saumon, Gilles Chabrier, and Armin Bergermann. We also acknowledge Cesare Cozza for providing initial data and support with the DFT calculations. H.X. and G.M. acknowledge financial support from the Swiss National Science Foundation (Grant No. PCEFP2\_203455). S.H. also acknowledges financial support from the Swiss National Science Foundation (Grant No. 200020\_215634).
\end{acknowledgments}

\bibliography{refs,bibdensehydrogen}

\setcounter{table}{0}
\renewcommand{\thetable}{S\arabic{table}}
\setcounter{figure}{0}
\renewcommand{\thefigure}{S\arabic{figure}}
\setcounter{equation}{0}
\renewcommand{\theequation}{S\arabic{equation}}

\appendix 

\clearpage 

\section*{Supplemental Material}

Consider a hydrogen system composed of $N$ protons and $N$ electrons in a periodic box of volume $L^3$. By introducing the dimensionless Wigner-Seitz parameter $r_s = (3/4 \pi N)^{1/3} L/a_0$, where $a_0$ is the Bohr radius, one can then write the mass density (in unit of $\textrm{g}/\textrm{cm}^3$) of the system as $\rho = 2.69467/r_s^3$. Let $(T_0, \rho_0), (T_1, \rho_1)$ be two state points on the temperature-density phase diagram, the corresponding Boltzmann distributions then read $p_i(\vec{x}) = e^{-\beta_i E_i(\vec{x})} / Z_i$ $(i=0, 1)$, respectively. $\vec{x} \equiv (\vec{r}_1, \dots, \vec{r}_N)$ denotes the proton coordinates, $\beta_i = 1/k_B T_i$, and $k_B$ is the Boltzmann constant. $Z_i = \int d\vec{x} e^{-\beta_i E_i(\vec{x})}$ is known as the partition function and directly related to the Helmholtz free energy $F_i = -\frac{1}{\beta_i} \ln Z_i$.

In this work, we use the flow matching method to benchmark the calculation of free energy difference (scaled by temperature) between the two states, $\beta_1 F_1 - \beta_0 F_0 = -\ln Z_1/Z_0$, against the conventional thermodynamic integration (TI) approach. Below we will discuss various aspects of the new method, including the underlying theoretical framework, training objectives, implementation details, and some illustrative examples of the results.

\section{Targeted free energy perturbation\label{sec: TFEP}}
In contrast to TI, flow matching is built on the framework of targeted free energy perturbation (TFEP)~\cite{PhysRevE.65.046122,PhysRevE.79.011113}, as already mentioned in the main text. Compared to the original free energy perturbation based on importance sampling~\cite{10.1063/1.1740409}, TFEP further leverages an invertible map in the configuration space to significantly increase state overlap and thereby reduces the variance of relevant estimators.

More specifically, consider a map $f$ that transforms coordinate samples $\vec{x} \sim p_0(\vec{x})$ of the state 0 into a new state $0^\prime$, whose probability density function is given by the change-of-variables formula:
\begin{align}
    p_0^\prime(\vec{x}) &= p_0(f^{-1}(\vec{x})) \left| \det\left( \frac{\partial f^{-1}(\vec{x})}{\partial \vec{x}} \right) \right| \nonumber \\
    &= \frac{1}{Z_0} e^{-\beta_0 E_0(f^{-1}(\vec{x}))} \left| \det\left( \frac{\partial f^{-1}(\vec{x})}{\partial \vec{x}} \right) \right| \nonumber \\
    &= \frac{1}{Z_0} e^{-\beta_0 E_0^\prime(\vec{x})},
\end{align}
where the transformed ``energy function" is defined as
\begin{equation}
    E_0^\prime(\vec{x}) = E_0(f^{-1}(\vec{x})) - \frac{1}{\beta_0} \ln \left| \det\left( \frac{\partial f^{-1}(\vec{x})}{\partial \vec{x}} \right) \right|.
\end{equation}
Similarly, the inverse of $f$ would transform the state 1 to a new state $1^\prime$ with energy function
\begin{equation}
    E_1^\prime(\vec{x}) = E_1(f(\vec{x})) - \frac{1}{\beta_1} \ln \left| \det\left( \frac{\partial f(\vec{x})}{\partial \vec{x}} \right) \right|.
\end{equation}
In practice, one would like to find an appropriate $f$ such that the transformed state $0^\prime (1^\prime)$ has significant overlap with the target state $1 (0)$. Based on the idea of importance sampling, the free energy difference can then be efficiently estimated as follows:
\begin{align}
    \beta_1 F_1 - \beta_0 F_0 &= -\ln \frac{Z_1}{Z_0} \nonumber \\
    &= -\ln \mathop{\mathbb{E}}_{\vec{x} \sim p_0(\vec{x})} \left[ e^{-\Phi_
    \rightarrow(\vec{x})} \right]
    = \ln \mathop{\mathbb{E}}_{\vec{x} \sim p_1(\vec{x})} \left[ e^{\Phi_\leftarrow(\vec{x})} \right],
    \label{Seq: TFEP}
\end{align}
where
\begin{subequations}
    \begin{align}
        \Phi_\rightarrow(\vec{x}) &= \beta_1 E_1^\prime(\vec{x}) - \beta_0 E_0(\vec{x}) \nonumber \\
    &= \beta_1 E_1(f(\vec{x})) - \beta_0 E_0(\vec{x}) - \ln \left| \det\left( \frac{\partial f(\vec{x})}{\partial \vec{x}} \right) \right|, \\
        \Phi_\leftarrow(\vec{x}) &= \beta_1 E_1(\vec{x}) - \beta_0 E_0^\prime(\vec{x}) \nonumber \\
    &= \beta_1 E_1(\vec{x}) - \beta_0 E_0(f^{-1}(\vec{x})) + \ln \left| \det\left( \frac{\partial f^{-1}(\vec{x})}{\partial \vec{x}} \right) \right|
    \end{align}
    \label{Seq: forward and reverse work}
\end{subequations}
are known as the \emph{forward} and \emph{reverse} work, respectively. These results are precisely the Eqs. (\ref{eq: TFEP}) and (\ref{eq: forward and reverse work}) in the main text.

Note that due to the limited number of samples in practice, the two estimators in \Eq{Seq: TFEP} are actually biased, which can be readily demonstrated using Jensen's inequality~\cite{PhysRevE.79.011113}. Specifically, the estimator involving the forward (reverse) work turns out to be an upper (lower) bound of the true free energy difference value. The extent of bias will, of course, depend on the quality of the map $f$. In this work, we found these two bounds are generally quite tight; see the sections below for some illustrative examples. As a result, we choose to simply perform an arithmetic mean of the two bounds to generate the final free energy and entropy data, although some more sophisticated and potentially superior approaches may also be considered~\cite{BENNETT1976245}. Another observation is that the estimation of \emph{absolute} entropy appears to be more straightforward within the TFEP framework compared to TI: once one of the two states is chosen to have known entropy, the practical implementation remains almost unchanged.

\section{Flow matching}
As mentioned in the main text, one appealing feature of the TFEP estimators in \Eq{Seq: TFEP} is that they depend solely on samples from the two end states, as opposed to TI which requires information along an entire path. The price one pays, however, is the design of a suitable map $f$, which is often challenging and lack of generic physical guidelines. This is arguably the main reason why the TFEP framework has not been widely used in realistic problems since its introduction.

The recent progress in the field of deep generative modeling has largely changed the situation. Specifically, the desired invertible map can be efficiently represented by a class of generative models called normalizing flow~\cite{Papamakarios2021}, making TFEP reemerge as a promising tool for free energy calculation. In previous attempts~\cite{10.1063/5.0018903, Wirnsberger_2022, Ahmad2021a, Molina-Taborda2024}, the training objectives have been exclusively based on variants of the Kullback-Leibler (KL) divergence, which provide direct access to the free energy difference as a variational upper or lower bound. This \emph{variational} training scheme has been successfully applied to a broad range of physical problems, including lattice models~\cite{PhysRevLett.121.260601,PhysRevLett.122.080602}, atomic solids~\cite{Wirnsberger_2022,Ahmad2021a}, quantum dots~\cite{JML-1-38}, uniform electron gases~\cite{10.21468/SciPostPhys.14.6.154}, and also dense hydrogen~\cite{PhysRevLett.131.126501}. However, it still has some drawbacks for the present purpose. First, the training process typically requires a large number of energy calculations of the underlying system, which would be time-consuming for \emph{ab initio} methods like DFT. As a result, many works have used empirical~\cite{10.1063/5.0018903,Wirnsberger_2022,Ahmad2021a} or machine-learned potentials~\cite{Molina-Taborda2024}, which would inevitably introduce new source of errors. Second, symmetry constraints often play an essential role in the efficient training of flow model, especially for homogeneous systems like pure hydrogen. Although several neural network designs have been proposed to meet various symmetry requirements~\cite{10.1063/5.0018903, Wirnsberger_2022}, they are still not satisfactory enough regarding a proper balance between the expressive power and computational cost. The continuous normalizing flow (CNF)~\cite{10.5555/3327757.3327764, zhang2018mongeampre} is arguably the most convenient and flexible candidate now to incorporate symmetries, but its variational training requires expensive numerical integrations of ordinary differential equation (ODE) and scales unfavorably for large system sizes.

The recently developed flow matching method~\cite{albergo2023building, lipman2023flow, liu2023flow} has largely overcome these difficulties. More specifically, one can smoothly connect the two end probability distributions $p_0(\vec{x}), \, p_1(\vec{x})$ by introducing an interpolant $I_t(\vec{x}_0, \vec{x}_1)$, where $t \in [0, 1]$ is a continuous ``time" parameter, such as
\begin{equation}
    I_{t=0}(\vec{x}_0, \vec{x}_1) = \vec{x}_0, \quad I_{t=1}(\vec{x}_0, \vec{x}_1) = \vec{x}_1.
\end{equation}
The essence is that the velocity field $\vec{v}_t(\vec{x})$ underlying such an interpolated ``probability path" can then be obtained by minimization of the following objective:
\begin{equation}
    \mathcal{L} = \mathop{\mathbb{E}}_{t \sim [0, 1]} \mathop{\mathbb{E}}_{\vec{x}_0 \sim p_0(\vec{x}_0)} \mathop{\mathbb{E}}_{\vec{x}_1 \sim p_1(\vec{x}_1)}
    \left|
    \vec{v}_t(I_t(\vec{x}_0, \vec{x}_1)) - \partial_t I_t(\vec{x}_0, \vec{x}_1))
    \right|^2.
    \label{eq: flow matching objective}
\end{equation}
Notice this expression requires only samples from the two end distributions; there are not any on-the-fly evaluations of configurational energy or ODE integration during the training stage. On the other hand, these types of costly computations are indeed present when estimating the free energy difference after the training is finished, as shown in Eqs.~(\ref{Seq: TFEP}) and (\ref{Seq: forward and reverse work}). In particular, the transformed coordinates $f(\vec{x})$ and corresponding log-Jacobian determinant $\ln \left| \det (\partial f / \partial \vec{x}) \right| \equiv \ln J_f(\vec{x})$ should be evaluated by jointly integrating the following set of ODEs~\cite{10.5555/3327757.3327764, zhang2018mongeampre}:
\begin{equation}
    \frac{d\vec{x}}{dt} = \vec{v}_t(\vec{x}), \quad \frac{d \ln J_f}{dt} = \nabla\cdot\vec{v_t(x)}.
    \label{eq: ODE}
\end{equation}
However, it should be noted that such computations are performed only once and thus still practically affordable.

As mentioned above, compared to previous network designs for discrete flow model~\cite{10.1063/5.0018903, Wirnsberger_2022}, the incorporation of various symmetries can also be more convenient for continuous normalizing flow. Specifically, the invariance of probability distribution can simply be achieved by requiring the underlying velocity field to be \emph{equivariant} under the desired symmetry operation $\mathcal{P}$:
\begin{equation}
    \vec{v}_t(\mathcal{P} \vec{x}) = \mathcal{P} \vec{v}_t(\vec{x}).
\end{equation}
To do this in practice, we can leverage many recent advances in various areas, such as natural language processing~\cite{NIPS2017_3f5ee243}, molecular simulation~\cite{khler2019equivariant, 10.1063/5.0018903}, graph neural networks~\cite{pmlr-v139-satorras21a}, and quantum many body computations~\cite{PhysRevB.91.115106, PhysRevResearch.2.033429, 10.21468/SciPostPhys.14.6.154, PhysRevResearch.4.023138, PhysRevLett.131.126501, PhysRevB.107.235139, PhysRevLett.130.036401, PhysRevB.110.035108}.

\section{Implementation details}
The MD generation of coordinate samples used for optimizing \Eq{eq: flow matching objective} and the energy computation appearing in the free energy estimators \Eq{Seq: TFEP} are both performed in the DFT level using the \texttt{QUANTUM ESPRESSO} code~\cite{QE-2009}. We use the exchange-correlation functional of Perdew, Burke and Ernzerhof (PBE)~\cite{PhysRevLett.77.3865} and the pseudopotential of projector-augmented wave~\cite{PhysRevB.50.17953} type. We choose the system size $N=128$, and the $\vec{k}$-space integration is performed on a $3 \times 3 \times 3$ Monkhorst-Pack grid~\cite{PhysRevB.13.5188}. The energy cutoff for the plane wave basis set and charge densities are $80$ and $800$ Ry, respectively. Note the parameters above have been checked to yield well converged pressures and energies. For MD simulations, the ion temperature is controlled using stochastic-velocity rescaling~\cite{10.1063/1.2408420} and the time step is set to 12 a.u.. 

The network we use for the velocity field $\vec{v}_t(\vec{x})$ is adapted from \texttt{FermiNet}~\cite{PhysRevResearch.2.033429}, which is permutation and translation equivariant by construction.  We also modify the pair distance features to comply with the periodic nature of the simulation box~\cite{10.21468/SciPostPhys.14.6.154, PhysRevResearch.4.023138, PhysRevB.107.235139, PhysRevLett.130.036401}. We choose a simple linear interpolating function
\begin{equation}
    I_t(\vec{x}_0, \vec{x}_1) = (1-t) \vec{x}_0 + t \vec{x}_1,
\end{equation}
with the caveat in mind that we are working on a periodic box. This setting turns out to be equivalent to the Riemannian flow matching on a torus as proposed in Ref.~\cite{chen2024flow}. To allow for more efficient use of the training data, we choose to further incorporate permutation symmetry explicitly into the flow matching objective \Eq{eq: flow matching objective} itself~\cite{klein2023equivariant}. In practice, this can be achieved by finding the permutation with minimal distance for each pair $\vec{x}_0, \vec{x}_1$ of samples using the Hungarian algorithm~\cite{10.1002/nav.3800020109}. 


\section{Illustrative examples of the results}
\begin{figure}[t]
    \centering
    \includegraphics[width=\columnwidth]{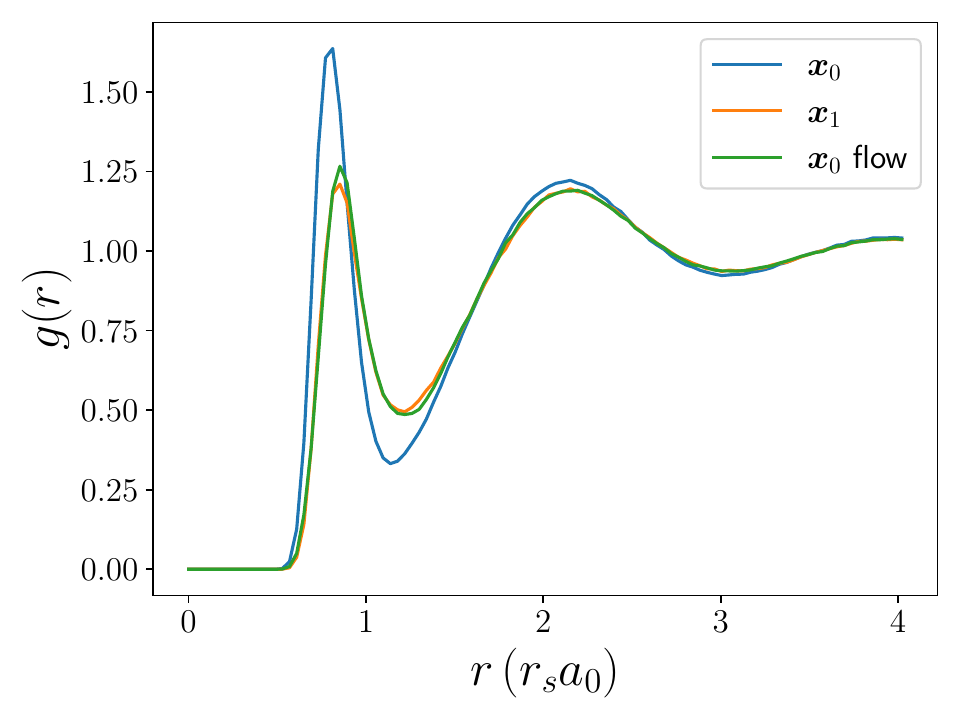}
    \caption{Radial distribution functions of the two states at $T_0 = 3000\textrm{K}$, $\rho_0 = 0.5\textrm{g}/\textrm{cm}^3$ and $T_1 = 3000\textrm{K}$, $\rho_1 = 0.55\textrm{g}/\textrm{cm}^3$, respectively. The line labeled with ``$\vec{x}_0$ flow" represents the new state obtained by transforming samples of state 0 using the velocity field trained with flow matching.
    }
    \label{fig: flow}
\end{figure}

To showcase the utility of TFEP and flow matching in practice, we consider as a typical example the two end states with $T_0 = 3000\textrm{K}$, $\rho_0 = 0.5\textrm{g}/\textrm{cm}^3$ and $T_1 = 3000\textrm{K}$, $\rho_1 = 0.55\textrm{g}/\textrm{cm}^3$, respectively. These states are near the molecular dissociation where the radial distribution function changes dramatically, as shown in Figure~\ref{fig: flow}. Note we have measured the distance in terms of $r_s a_0$; under such a scale, the simulation box size is $(4 \pi N/3)^{1/3}$ and does not depend on the density of the system. This would facilitate the interpolation between the two Boltzmann distributions as done in the flow matching method. Fig.~\ref{fig: flow} also shows the radial distribution function of the new state transformed from state 0, obtained by integrating the first ODE in \Eq{eq: ODE} using the trained velocity field. The result is clearly very close to the target state 1, indicating a significant state overlap.

\begin{figure}[t]
    \centering
    \includegraphics[width=\columnwidth]{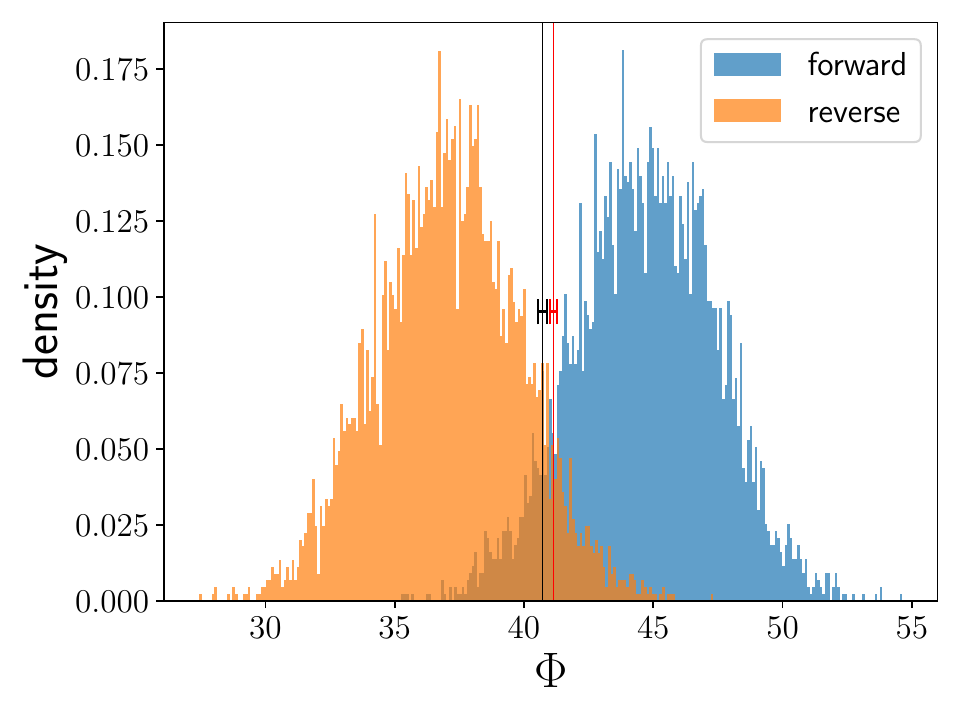}
    \caption{Histograms for the distribution of forward and reverse work, defined in \Eq{Seq: forward and reverse work}, between the two states at $T_0 = 3000\textrm{K}$, $\rho_0 = 0.5\textrm{g}/\textrm{cm}^3$ and $T_1 = 3000\textrm{K}$, $\rho_1 = 0.55\textrm{g}/\textrm{cm}^3$. The vertical red (black) line is the estimated upper (lower) bound for free energy difference; see \Eq{Seq: TFEP} and the discussions therein.
    }
    \label{fig: work}
\end{figure}

After the training is complete, one can further use \Eq{Seq: TFEP} to estimate the upper and lower bound of free energy difference between the two states, as shown by the red and black vertical line in \Fig{fig: work}, respectively. The distance between these two bounds is roughly within their own error bars, which implies our estimation have been tight enough. Figure~\ref{fig: work} also shows histograms for the distribution of forward and reverse work $\Phi_\rightarrow$, $\Phi_\leftarrow$ as defined in \Eq{Seq: forward and reverse work}. The significant overlap between these two work distributions indicates the high quality of the trained velocity field in effectively connecting the two end states being considered. Furthermore, one can clearly see the estimated free energy difference values coincide well with the intersection point of the two work distributions, where the probability densities are equal. This is a direct consequence of the fluctuation theorem as derived in Ref.~\cite{PhysRevE.79.011113}.

\begin{figure}[t]
    \centering
    \includegraphics[width=\columnwidth]{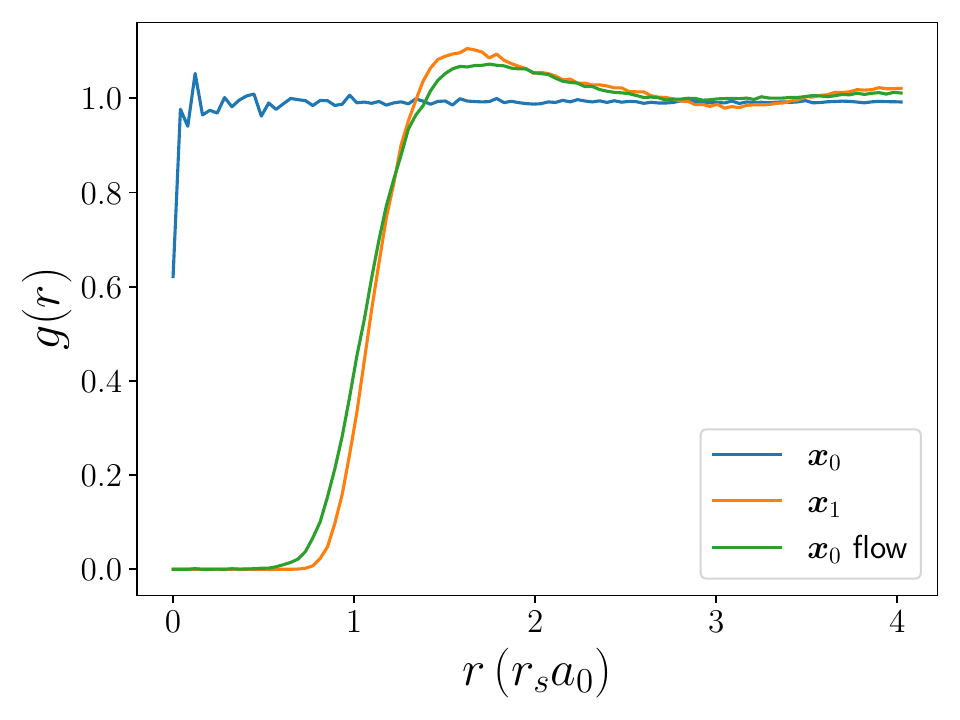}
    \caption{Same as Fig.~\ref{fig: flow}, except the states 0 and 1 correspond to the uniform distribution and the reference state at $T = 5000\textrm{K}$, $\rho = 1.4\textrm{g}/\textrm{cm}^3$, respectively.
    }
    \label{fig: flow_uniform}
\end{figure}

\begin{figure}[t]
    \centering
    \includegraphics[width=\columnwidth]{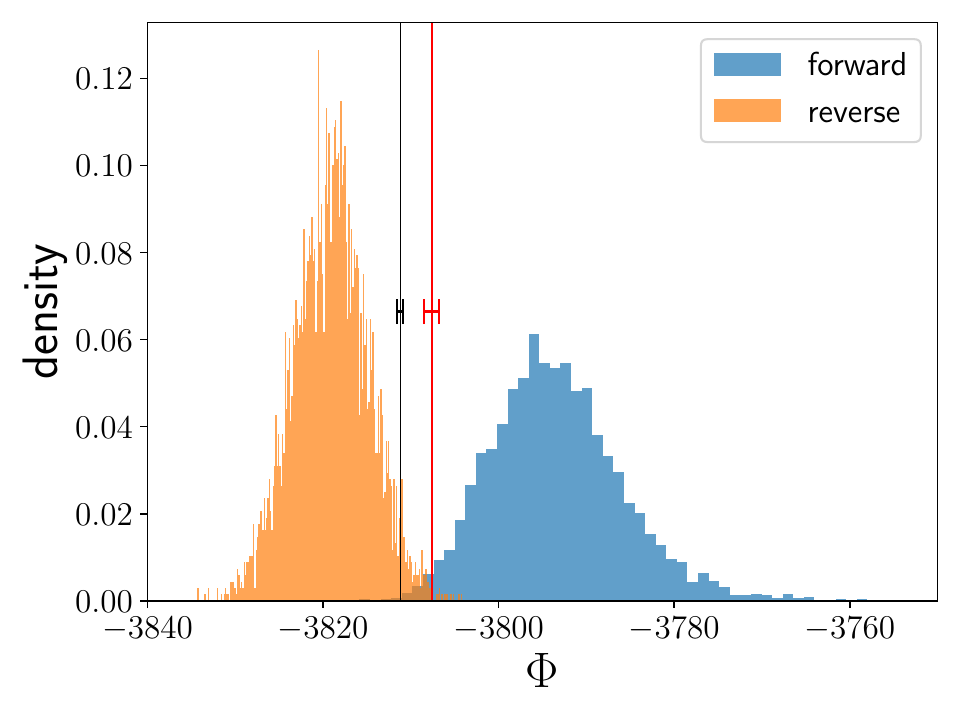}
    \caption{Same as Fig.~\ref{fig: work}, except the states 0 and 1 correspond to the uniform distribution and the reference state at $T = 5000\textrm{K}$, $\rho = 1.4\textrm{g}/\textrm{cm}^3$, respectively.
    }
    \label{fig: work_uniform}
\end{figure}

We also use the same workflow outlined above to calculate the \emph{absolute} free energy and entropy of the reference state at $T = 5000\textrm{K}$ and $\rho = 1.4\textrm{g}/\textrm{cm}^3$, as mentioned in the main text. In practice, we choose the ``tractable state" to correspond to a simple uniform distribution. The resulting radial distribution functions and free energy bounds are shown in Figures~\ref{fig: flow_uniform} and \ref{fig: work_uniform}, respectively. Notice that in the present case, the two end states differ more substantially than those in \Fig{fig: flow}, making the flow matching training of the velocity field more challenging. As a result, the forward and reverse work distributions also have less overlap, and the upper and lower bounds for the free energy difference are less tight compared to \Fig{fig: work}. Nonetheless, the absolute entropy of the reference state can still be estimated with a relative error of $0.2\%$, which is comparable to the results reported by Morales \emph{et al.}~\cite{PhysRevE.81.021202} using Hamiltonian thermodynamic integration and already satisfactory for the present purpose (see \Fig{fig: entropy_pressure}(a) in the main text).

\section{Effect of thermodynamic inconsistency on entropy calculation}
\begin{figure}[t]
    \centering
    \includegraphics[width=\columnwidth]{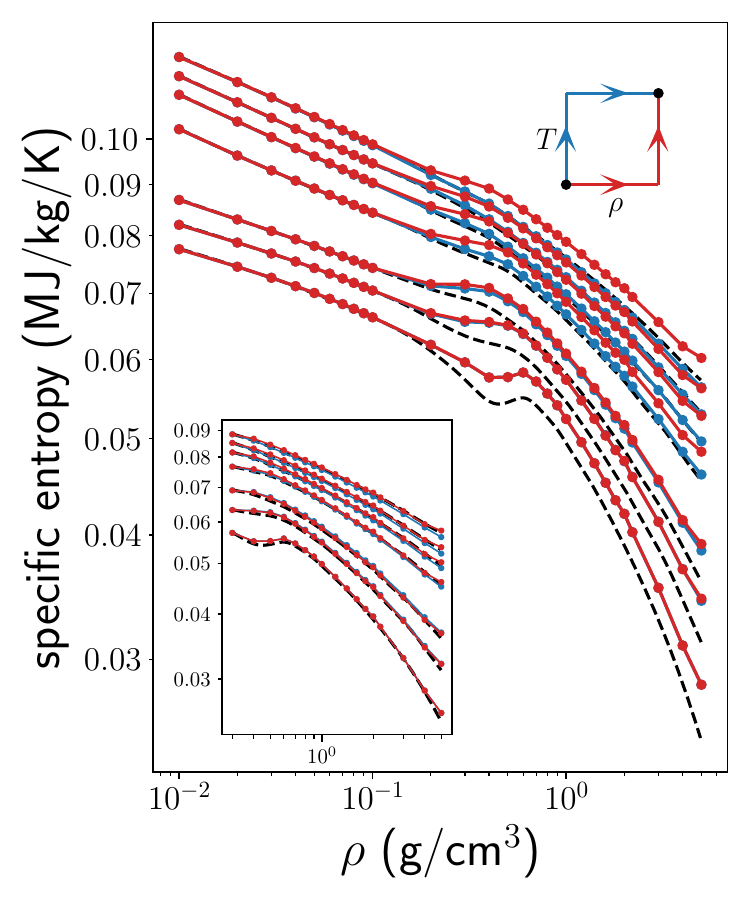}
    \caption{Several specific entropy isotherms of REOS3 computed by TI over a wide density range ($0.01 \leqslant \rho \leqslant 5\textrm{g}/\textrm{cm}^3$, main panel) and only the \emph{ab initio} region at high densities ($\rho \geqslant 0.3 \textrm{g}/\textrm{cm}^3$, inset). The corresponding temperatures for the isotherms are (from bottom to top) 3000, 4000, 5000, 8000, 10000, 12000, and 15000\textrm{K}. The red (blue) solid lines correspond to the results of integrating along isotherm (isochore) first, as schematically illustrated on the top-right corner. For both regions, the reference point is set to the one with lowest temperature and density. The dashed lines are the entropy data produced by Miguel \emph{et al.}~\cite{Miguel2016}.
    }
    \label{fig: entropy_TI_REOS3}
\end{figure}

In \Fig{fig: flux_REOS3} of the main text, we have shown the entropy calculated by a single TI over the entire phase region of REOS3 can be thermodynamically inconsistent in the intermediate region between different theories. By comparing the maximum value $10^{-3}\textrm{MJ/kg/K}$ of the color bar in \Fig{fig: flux_REOS3} to the scale of \Fig{fig: entropy_pressure}(a) in the main text, one can conclude that a local inconsistency of such a magnitude will likely lead to significant discrepancy of the entropy. To illustrate this, we compute the REOS entropy by integrating either along the isotherm or isochore first, as shown by the red and blue solid lines in Figure~\ref{fig: entropy_TI_REOS3}, respectively. Note for a given grid of points, we have selected the one with lowest temperature and density as the reference point for integration. One can see for a large phase region (the main panel of \Fig{fig: entropy_TI_REOS3}) that covers both the chemical model and \emph{ab initio} data, the results from the two integration paths begin to diverge significantly at intermediate densities, which is a direct consequence of the inconsistency observed in \Fig{fig: flux_REOS3} of the main text. On the other hand, such a divergence becomes nearly negligible when the integration is performed exclusively over the \emph{ab initio} region, as shown in the inset of \Fig{fig: entropy_TI_REOS3}.

\begin{figure}[t]
    \centering
    \includegraphics[width=\columnwidth]{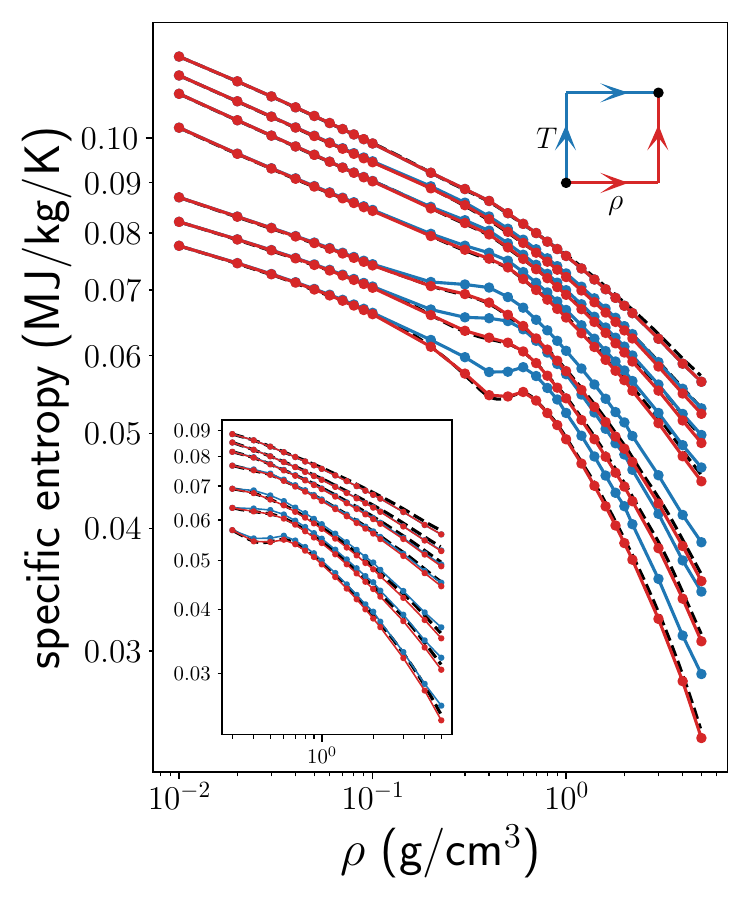}
    \caption{Same as \Fig{fig: entropy_TI_REOS3}, except the reference point (for TI calculation of the solid lines) is set to be at $T = 15000\textrm{K}$ and the lowest density.
    }
    \label{fig: entropy_TI_REOS3_Tanchor_15000}
\end{figure}

The subtleties involved in the implementation of TI arise not only from different choices of integration path, but also the reference point. In fact, the latter can be seen as an alternative manifestation of the former, which can be understood by noting that an integration path can belong to different types (i.e., isotherm or isochore first) when viewed by different reference points. As an illustration, \Fig{fig: entropy_TI_REOS3} also shows the entropy reported by Miguel \emph{et al.}~\cite{Miguel2016} (dashed lines). According to the authors, they processed the original REOS3 data by integrating along isotherm first, hence in accordance with the red solid lines of our result. However, these two sets of entropies clearly differ by an offset in the \emph{ab initio} region of the main panel. The remedy of this apparent disagreement turns out to be surprisingly easy: notice all the entropy isochores already align well, we just need to locate the reference point on another ``correct" isotherm, say, 15000K. After doing this, the red solid lines then coincide perfectly with the dashed lines from Ref.~\cite{Miguel2016}, as shown in Figure~\ref{fig: entropy_TI_REOS3_Tanchor_15000}.

Our reproduction of the entropy results by Miguel \emph{et al.}~\cite{Miguel2016} based on an \emph{ad hoc} choice of the integration path and reference point, as done above, demonstrates how fragile the ``global TI" approach can be for an EOS with significant thermodynamic inconsistencies. Notice the solid and dashed lines in \Fig{fig: entropy_TI_REOS3} differ by roughly $10\%$, which can make a big difference on planetary model predictions as shown in the main text.

\section{More details on the thermodynamically consistent construction of our EOS}
To construct an EOS that is thermodynamically consistent across several distinct theories, such as \emph{ab initio} DFT and SCvH, one needs to first calculate the free energy and entropy of each region \emph{separately} using the standard TI approach. We then perform a two-dimensional spline interpolation over each set of free energy data, following a similar approach to Militzer and Hubbard~\cite{Militzer_2013}. This in principle allows us to reproduce the original pressure and energy data by performing partial derivatives. The quality of such reproduction is, of course, closely related to the original level of thermodynamic consistency in each region.
Once the free energy interpolation for each separate region is ready, we can then glue them together by performing another spline interpolation along their boundaries. Note one can utilize relevant partial derivative values derived from the free energy interpolations on both sides to ensure a smoother connection.

In this work, we attempted an interpolation between the \emph{ab initio} DFT-PBE data at $\rho \geqslant 0.3\textrm{g}/\textrm{cm}^3$ and the SCvH EOS at $\rho \leqslant 0.1\textrm{g}/\textrm{cm}^3$. Note we have chosen the locations of the density boundaries based on reasonable assumptions about the range of validity of both theories. Intuitively speaking, the size of the intermediate density gap reflects a trade-off between our confidence in available (but potentially conflicting) data and the flexibility to perform interpolations between them.


\begin{figure*}[t]
    \centering
    \includegraphics[width=0.9\textwidth]{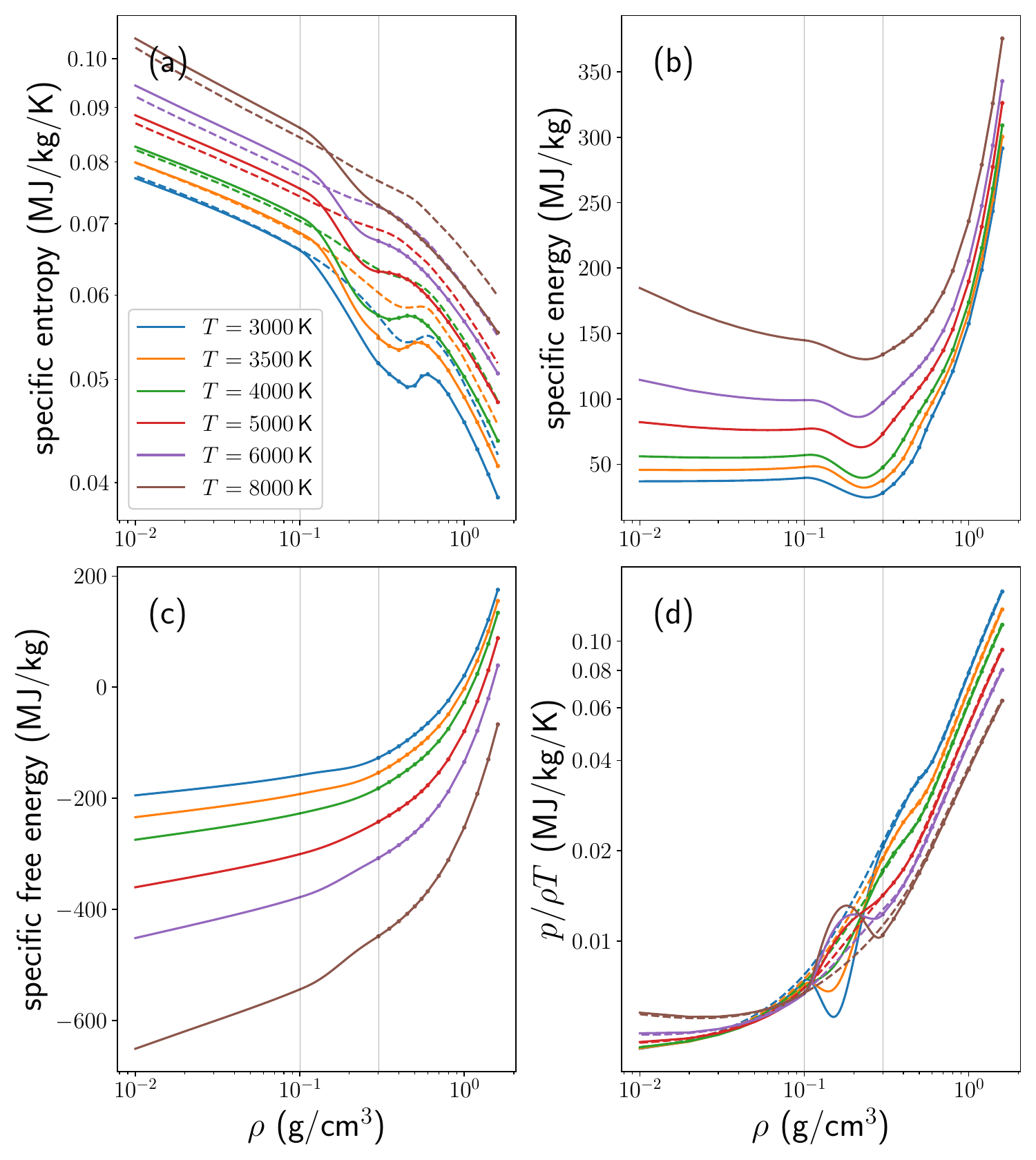}
    \caption{The hydrogen EOS (solid lines) constructed in the same way as \Fig{fig: final_eos} in the main text, except that we adopt the \emph{original} absolute entropy of the SCvH EOS without subtracting the constant $k_B \ln 2 / m_\textrm{p} = 0.0057 \, \textrm{MJ/kg/K}$. The dashed lines are the original entropy and pressure isotherms of Miguel \emph{et al.}~\cite{Miguel2016}, obtained by performing a single TI over the entire phase region of REOS3.
    }
    \label{fig: final_eos_wrong_SCvH}
\end{figure*}

In principle, the free energy interpolation would be perfectly smooth if the absolute entropy and energy (measured relative to the same baseline) were exact on both sides. However, this is not the case in practice. First, it is generally believed that only the \emph{relative energies} produced by DFT (using any given exchange-correlation functional) are reliable, rather than the absolute ones. This indicates that we are allowed to globally shift the DFT energies as we want, which, of course, leaves the entropy (and any other thermodynamic properties) unchanged. Second, note that in the DFT-MD region, we have implicitly neglected the entropy contribution $k_B \ln 2 / m_\textrm{p} = 0.0057 \, \textrm{MJ/kg/K}$ from the proton spin degrees of freedom, as they are irrelevant for essentially classical protons at high temperatures. However, this constant plays an important role when trying to connect with SCvH, which has taken account of accurate energy levels arising from both proton spin isomers (i.e., para- and ortho-hydrogen) when constructing the partition function of molecular hydrogen gas at low densities~\cite{PhysRevA.44.5122}. As a result, the constant must be subtracted from the original SCvH data to ensure a truly fair matching of the absolute entropies on both sides. This is precisely what we have done to produce the final EOS shown in \Fig{fig: final_eos} of the main text.

To get a feeling on how \emph{good} is the interpolation quality, it is instructive and interesting to see how \emph{bad} it will become if we just mindlessly adopt the \emph{original} SCvH entropy data without subtracting the constant mentioned above. The resulting EOS based on the same interpolation procedure is shown in Figure~\ref{fig: final_eos_wrong_SCvH}. Notice that although we have selected a DFT energy offset to connect the free energy isotherms on both sides as smoothly as possible (see panel (c)), there still remains a noticeable mismatch. In fact, according to the basic relation $F = E - TS$, the free energy isotherms of SCvH are apparently \emph{wider} than the \emph{ab initio} region due to \emph{too large} entropy on the SCvH side. On the other hand, \Fig{fig: final_eos_wrong_SCvH}(d) shows the interpolated pressure is also problematic. In particular, the wild wiggling behavior at the intermediate densities can even cause violation of the basic mechanical stability constraint $(\partial p / \partial \rho)_T > 0$~\cite{1995ApJS...99..713S} and hence definitely \emph{unphysical}.


The observations above indicate that \emph{the interpolation quality serves as a very sensitive and stringent “detector” of the accuracy of both theories}. In particular, we have been \emph{forced} to identify, confirm and ``fix" the constant shift in original SCvH entropy data, \emph{even if we were totally unaware of this subtlety in the early stages of this work}. This is a very compelling demonstration of the power of our new EOS construction protocol. In particular, such an iteratively ``\emph{self-correcting}" mechanism possessed by the protocol is exactly what we need to ensure steady convergence towards a conclusive hydrogen EOS for planetary modeling.

Although we have achieved a fairly excellent matching between SCvH and DFT-PBE-MD, there still remains some small errors. In particular, note we actually have also added an extra (and very small) entropy offset of $0.001 \, \textrm{MJ/kg/K}$ on the DFT side to make the interpolation shown in \Fig{fig: final_eos} of the main text look slightly better. In principle, such remaining errors can be eliminated by using more accurate chemical models at low densities or electron structure methods (such as QMC) at high densities. Besides, the nuclear quantum effect may also have a slight impact on the \emph{ab initio} entropy at low temperatures, which has been ignored in this work. This can be remedied either by adding some semi-empirical corrections~\cite{PhysRevB.87.014202, PhysRevB.88.045122} or resorting to more accurate (and also more costly) path integral MD simulations~\cite{PhysRevLett.110.065702}.

\section{More details of comparison between different Jupiter's adiabats}
\begin{figure}[t]
    \centering 
    \includegraphics[width=\columnwidth]{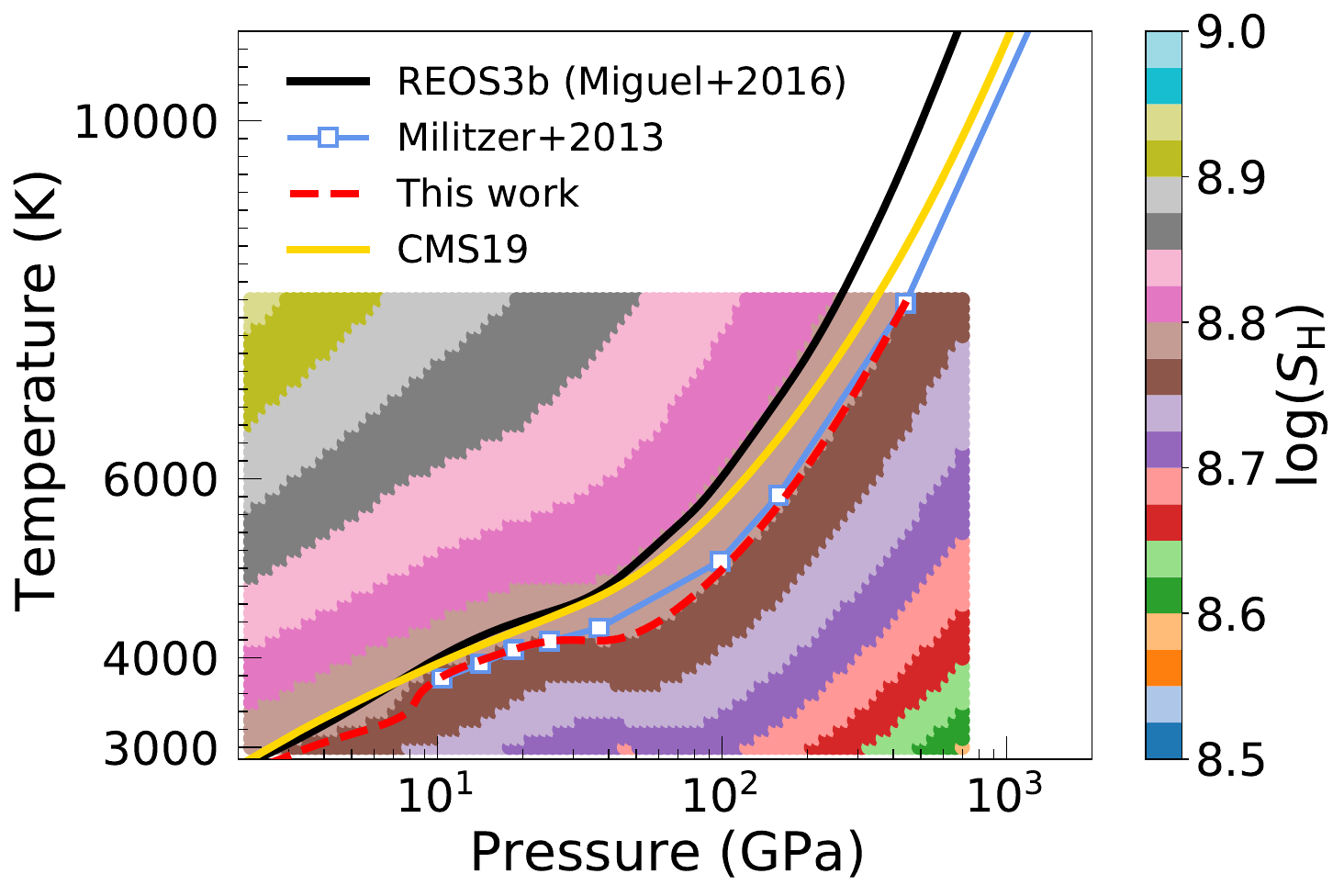}
    \caption{Jupiter adiabats obtained from different EOSs. Compared to \Fig{fig:saburo} of the main text, we also additionally show the adiabat derived from the CMS19 EOS~\cite{Chabrier_2019}, as well as the entropy of our EOS by colored regions.
    }
    \label{fig:saburo_supp}
\end{figure}

Figure~\ref{fig:saburo_supp} shows some more details of Jupiter's adiabats calculated using various hydrogen EOSs. Compared to \Fig{fig:saburo} of the main text, there is additionally the adiabat derived from the CMS19 EOS~\cite{Chabrier_2019} (yellow line), which is clearly \emph{smoother} in the \emph{ab initio} region than both our result (red dashed line) and MH13 (blue line). This difference originates from the significant, non-monotonic slope change of our entropy isotherms, as shown in \Fig{fig: entropy_pressure}(a) of the main text. In \Fig{fig:saburo_supp}, we also directly illustrate the entropy of our EOS by colored regions. The change of slope is again clearly evident, indicating that our EOS better captures the pressure-induced molecular dissociation than CMS19.

By definition, our calculated adiabat (red dashed line) follows the shape of a certain ``stripe" of constant entropy, as clearly shown in \Fig{fig:saburo_supp}. Note that the slightly imperfect behavior of our adiabat between 1 and 10GPa corresponds to the interpolation region of our EOS; see the previous section for more discussions. However, it is worth emphasizing that this remaining error will not affect the accuracy and reliability of our adiabat at higher pressures, where the disagreements between various adiabats make the biggest difference on resulting planetary model predictions~\cite{Miguel2016, howard2023}.

In \Fig{fig:saburo_supp}, our adiabat is confined to the \emph{ab initio} region explored in this work, specifically up to $T = 8000$K and $\rho = 1.6 \textrm{g}/\textrm{cm}^3$. It is in principle just a matter of additional DFT-MD simulations or interpolations with other theories~\cite{1995ApJS...99..713S, PhysRevE.58.4941, PhysRevLett.85.1890, PhysRevE.63.066404} to extend the hydrogen EOS to higher temperatures and densities. This will enable us to accurately study Jupiter's deeper interior beyond the red dashed line in \Fig{fig:saburo_supp}, or other astrophysical objects under even more extreme conditions, such as solar-type stars.

\end{document}